\documentclass[conference]{IEEEtran}
\IEEEoverridecommandlockouts
\usepackage{cite}
\usepackage{amsmath,amssymb,amsfonts}
\usepackage{algorithmic}
\usepackage{graphicx}
\usepackage{textcomp}
\usepackage[dvipsnames]{xcolor}

\usepackage{subcaption}

\usepackage[T1]{fontenc}

\newcommand{\lmtt}{\fontfamily{lmtt}\selectfont}

\usepackage{hyperref}
\usepackage{booktabs}

\usepackage{multirow}
\usepackage{stfloats}

\def\BibTeX{{\rm B\kern-.05em{\sc i\kern-.025em b}\kern-.08em
    T\kern-.1667em\lower.7ex\hbox{E}\kern-.125emX}}

\usepackage{soul}
\newcommand{\ctext}[3][RGB]{%
  \begingroup
  \definecolor{hlcolor}{#1}{#2}\sethlcolor{hlcolor}%
  \hl{#3}%
  \endgroup
}

\usepackage{listings}
\lstdefinelanguage{oml}
{
  sensitive=true,
  morecomment=[l]{--},
  morestring=[b]',
  classoffset=0,
  morekeywords={concept,@rdfs,@dc,relation,ci,from,to,reverse,asymmetric,irreflexive,forward, mission,components,objectives,base,aspect,restricts,some,rule},
  classoffset=1,
  morekeywords={======, MISSION,MISSIONS,VOCABULARY, DESCRIPTION}
}

\definecolor{commentgreen}{rgb}{0.0, 0.4, 0.1}
\definecolor{commentblue}{rgb}{0.0, 0.1, 0.4}
\definecolor{typegreen}{rgb}{0.0, 0.4, 0.0}
\definecolor{keywordred}{rgb}{0.4, 0.0, 0.1}

\definecolor{LightGray}{rgb}{0.97,0.97,0.97}

\lstdefinelanguage{SPARQL}{
  basicstyle=\small\lmtt,
  backgroundcolor=\color{LightGray},
  columns=fullflexible,
  breaklines=false,
  sensitive=true,
  frame=bt,
  aboveskip=1em,
  belowskip=1em,
  xleftmargin=.5em,
  xrightmargin=.5em,
  framexleftmargin=.5em,
  framextopmargin=.5em,
  framexbottommargin=.5em,
  framexrightmargin=.5em,
  tabsize = 2,
  showstringspaces=false,
  morecomment=[l][\color{gray}]{\#},       
  morecomment=[n][\color{blue}]{<http}{>}, 
  morestring=[b][\color{OliveGreen}]{\"},  
  keywordsprefix=?,
  classoffset=0,
  keywordstyle=\color{Sepia},
  morekeywords={},
  classoffset=1,
  keywordstyle=\color{Purple},
  morekeywords={rdf,rdfs,owl,xsd,purl},
  classoffset=2,
  keywordstyle=\color{MidnightBlue},
  morekeywords={
    SELECT,CONSTRUCT,DESCRIBE,ASK,WHERE,FROM,NAMED,PREFIX,BASE,OPTIONAL,
    FILTER,GRAPH,LIMIT,OFFSET,SERVICE,UNION,EXISTS,NOT,BINDINGS,MINUS,a
  }
}

\begin{document}

\title{Towards Understanding and Analyzing Rationale in Commit Messages using a Knowledge Graph Approach}

\author{\IEEEauthorblockN{Mouna Dhaouadi}
\IEEEauthorblockA{\textit{DIRO} \\
\textit{Universit\'{e} de Montr\'{e}al}\\
Montr\'{e}al, Canada \\
mouna.dhaouadi@umontreal.ca}
\and
\IEEEauthorblockN{Bentley James Oakes}
\IEEEauthorblockA{
\textit{DIRO}, \textit{Universit\'{e} de Montr\'{e}al}\\
\textit{GIGL, Polytechnique Montr\'{e}al}\\
Montr\'{e}al, Canada \\
bentley.oakes@polymtl.ca}
\and
\IEEEauthorblockN{Michalis Famelis}
\IEEEauthorblockA{\textit{DIRO} \\
\textit{Universit\'{e} de Montr\'{e}al}\\
Montr\'{e}al, Canada \\
famelis@iro.umontreal.ca}

\thanks{\textcolor{red}{Author pre-print. Accepted at MDEIntelligence 2023. Final published version may differ. \copyright 2023 IEEE.}}}

\maketitle

\thispagestyle{plain}
\pagestyle{plain}

\begin{abstract}
Extracting rationale information from commit messages allows developers to better understand a system and its past development. Here we present our ongoing work on the Kantara end-to-end rationale reconstruction pipeline to a) structure rationale information in an ontologically-based knowledge graph, b) extract and classify this information from commits, and c) produce analysis reports and visualizations for developers. We also present our work on creating a labelled dataset for our running example of the Out-of-Memory component of the Linux kernel. This dataset is used as ground truth for our evaluation of NLP classification techniques which show promising results, especially the multi-classification technique XGBoost.
\end{abstract}

\section{Introduction}
\label{sec:intro}

The development of software systems involves constant decision-making at various levels of abstraction and degrees of importance.
For every such decision, the system developer has a \emph{rationale}, i.e., the \textit{why} reasoning that explains their decision.
Rationale may be more or less explicitly articulated and is often left implicit or unrecorded.
Regardless, it is a very useful piece of information for the development team, as it can  be used to learn from mistakes or reuse solution patterns~\cite{burge2008rationale}.
Many researchers have thus attempted to extract rationale and to help developers leverage it~\cite{noble1988issue,conklin1988gibis,burge2005software,liang2012learning}.


In this article, we focus on the rationale expressed in developer commit messages submitted to version control systems such as Git.
Tian \textit{et al.} have found that a quality of a ``good'' commit message is that it contains rationale information~\cite{tian2022makes}.
Such information can help other developers understand and contextualize the proposed changes with respect to other project commits.
Crucially, it also allows other stakeholders to assess the \textit{quality} of the proposed changes and how well the changes fit with the project requirements.
Over time, these messages help build a shared understanding of the design and behavior of the software system among the development team.

The primary challenge to understanding rationale is that rationale information in commit messages is not typically expressed in a structured representation or model, and instead is embedded in natural language text. 
Therefore, identifying and understanding the rationale depends on deep prior knowledge of the project context.
This increases the mental effort required to understand a commit, its key decisions, and their rationale.
From the point of view of maintaining a shared understanding, this makes it \textit{harder to on-board new contributors}, and makes it \textit{difficult to assess the quality of the provided rationale of each new commit}.
Finally, it makes it difficult to \textit{establish and maintain traceability} with decisions and rationale information elsewhere in the project.





Here, we here present our results towards extracting models of rationale from commit messages using the Kantara framework for end-to-end rationale reconstruction~\cite{dhaouadi2022end}. Kantara defines an \textit{information inference component} to structure and extract rationale information, and an \textit{analysis interface} to understand the extracted rationale.
We apply a proof-of-concept demonstration of Kantara to a set of 180 commits from a Linux kernel subsystem. 
The main contribution is showcasing the practicality of end-to-end rationale extraction and analysis for a real world software project, and highlighting the challenges.


In Section~\ref{sec:example} we explain our running example of the OOM-Killer subsystem in the Linux kernal, our approach to labelling this subsystem's commits with the rationale information, and discuss our insights on how kernel developers explain their commits. 
The extraction of rationale information from commits is presented in Section~\ref{sec:extracting}, where Natural Language Processing (NLP) techniques are used to automatically classify commit sentences. From our ground truth dataset, we quantitatively evaluate the performance of our classifiers. Section~\ref{sec:structuring} provides our contribution on the structuring of this rationale information in an ontologically-based \textit{knowledge graph}~\cite{oakes2021structuring}, along with our efforts on representing this graph as an ontology in the openCAESAR framework~\cite{Elaasar2023} to leverage ontological semantics for inferencing. Section~\ref{sec:analysis} discusses the analysis of the rationale information once it is placed in the ontological format. We present a query operating on this graph to extract valuable information about commit sentences, and an interactive visualization allowing a user to better understand how rationale is presented in commits by different authors. In Section~\ref{sec:related_work}, we overview related work, while Section~\ref{sec:conclusion} concludes with a discussion on challenges we faced and the future work to address them.





\section{Running Example: Linux Subsystem}
\label{sec:example}

This section will discuss our running example of the \textit{Out-of-Memory} (OOM) memory management subsystem in the Linux kernel. First, the kernel itself and relevant development information is presented. Then, we discuss the OOM subsystem itself. Finally, we discuss our ongoing work on labelling rationale information in the commits of the OOM subsystem.

\subsection{Linux Kernel and Development}

Since its creation by Linus Torvalds in 1991, the Linux kernel has grown to run on devices ranging from mobile phones and tablets to supercomputers. This wide-spread adoption is partially due to its open-source and collaborative  development.

The main development channel for Linux is the Linux Kernel Mailing List (LKML)\footnote{Various archives exist for the LKML such as \url{https://lkml.org/}.}. The LKML contains email threads concerning every aspect of Linux development, including bug reports and potential patches. 
For example, this email\footnote{\url{https://lore.kernel.org/all/20100527180431.GP13035@uudg.org/}} is a request for comment (RFC) on a potential patch to be applied to the kernel. Once the patches are accepted by a \textit{subsystem maintainer}, they are passed through the maintainer hierarchy until the patches arrive at Linus Torvalds himself to be applied and released as a new version of Linux.


 To maintain a high degree of quality in the Linux kernel, there are requirements for patch submission\footnote{\url{https://www.kernel.org/doc/html/latest/process/submitting-patches.html}.}. Among these, the subject line for the email containing the patch must fit a particular standard. This patch email must also contain a few lines or paragraphs describing the motivation/rationale behind the patch, and the impact on the kernel the patch will have. Traceability information is provided in multiple ways, such as a) patches are encouraged to add explicit links to LKML discussions in the patch descriptions, and b) patches must have a \textit{summary phrase} (such as ``oom: give the dying task a higher priority'') which allows for tracing this patch across commits and LKML discussions. Linux kernel commit messages are thus a comprehensive repository of decision/rationale information to evaluate our approach.
 

\subsection{Out-of-Memory Killer (OOM-Killer) Subsystem}

The OOM-Killer kernel subsystem frees up memory when tasks have requested all available memory, preventing the system from crashing due to the lack of available memory. The OOM-Killer performs two primary tasks:  a) it selects a task to kill, then b) it forces that task (the \textit{OOM victim}) to release its memory and exit. The selection process is particularly relevant for our examination, as there are rich discussions about the best way to select the task to kill. For example, one code change was to reduce the chance of selecting kernel tasks specifically,  later reversed in favor of a uniform selection process.

This OOM approach has been controversial since its origin in 1998, as some developers reason that the system should be allowed to crash in the presence of faulty software, or express concern that desirable user processes (such as a windowing environment) may be selected for killing.
 
\subsection{Labelling Dataset for OOM Commits}
\label{sec:oom_labelling}

\begin{table}[t]
\small
  \caption{Codebook}
  \label{tab:codebook}
  \begin{tabular}{p{2.2cm}p{5.8cm}}
    \toprule
    Label & Meaning\\
    \midrule
    Decision& An action or a change that has been made, including a description of the patch behaviour \\
     Rationale& Reason for a decision or value judgment
     \\
   Supporting Facts & A narration of facts used to support a  decision \\
     Inapplicable &   Pre-processing error or bad sentences\\
     & (i.e., does not contain English sentences) \\
  \bottomrule
\end{tabular}

\end{table}

To evaluate our Kantara framework (Section~\ref{sec:extracting}), we require a form of ground truth to measure our extraction approaches. For this purpose, we are systematically creating a dataset of labelled commit messages for the OOM subsystem indicating the rationale information present in each commit sentence~\cite{FSE-SRC}.


\subsubsection{Commit Pre-processing}

To create our data set, 
we selected 180 commits from the commit history
of the OOM-killer file\footnote{\url{https://github.com/torvalds/linux/commits/master/mm/oom\_kill.c} accessed on 2023-01-12}. 
We did not include any merge commits, i.e., commits whose messages start with \textit{Merge tag}.
We pre-processed the messages of these commits: For each message, we removed the meta-data at the end of the message (e.g, information like \textit{Signed-off-by}  and \textit{Suggested-by} were removed). We also removed URLs, references to other resources, and call traces using regular expressions. Afterwards, we split the message into sentences and kept only sentences with more than 3 characters and that are not source code. We identified the source code in the message body using heuristics like identifying keywords or symbols such as \textit{git}, \textit{\$cd} or \textit{\$echo}. These keywords came from manually investigating the data. 

\subsubsection{Sentence Labelling Procedure}

We performed six iterations of piloting rounds and consolidation meetings during which the three annotators (a PhD student, a post-doctoral researcher, and a professor) considered 38 randomly-chosen commits in total (which they annotated independently).  Finally, we reached a consensus regarding the set of labels to use and our understanding of each label as shown in Table~\ref{tab:codebook}.


We conducted the labelling by batches where if two annotators said \textit{yes} to a label, then this was taken as the consensus. During the labelling process, Fleiss Kappa averaged 0.69 for  eight rounds (so far).  This indicates strong agreement considering the subjective nature of rationale~\cite{burge2008rationale}. 


\subsubsection{Sentence Classification}

Our rounds of labelling and consensus-building led us to further insights on how to classify each sentence. Consequently, we have updated our representation of rationale from what we had previously reported in our earlier work~\cite{dhaouadi2022end}. 
In particular, as we now examine commits at a \textit{sentence-level}, each sentence is given multiple classifications. We have also introduced the concept of a \textit{supporting fact} to indicate sentences which do not themselves contain rationale, but instead present the current state of the system  as explanatory text for the commit.

That is, a supporting fact is information in a sentence where a developer discusses the currently existing state of the system, at the moment before they propose a change. An example would be the description of the behaviour of a previous commit, such as the third sentence in Table~\ref{tab:coloured_commit}. Due to this reference to the past or current state of the system, these supporting facts are often found in the first half of the commit, and before the sentences containing rationale. In contrast, a \textit{decision} provides information about the state of the system \textit{after} the patch is applied, and thus it refers to the system's \textit{future} state. \textit{Rationale} is the \textit{reason} for \textit{why} a decision is taken, such as a \textit{value judgement} about undesirable behavior.

Table~\ref{tab:coloured_commit} reproduces a commit from our dataset, along with a color-coded multi-label classification for each sentence. As an example of the labelling, the first sentence (the \textit{summary phrase} of the commit) is labelled as a \textit{decision} as it states the patch's change. The fourth sentence contains \textit{rationale} as a value judgement (``something stupid'') and \textit{supporting facts} (``only takes 33+ privileged tasks'').



\begin{table*}[]
    \centering
    \begin{tabular}{p{14cm}|c}
       \textbf{Sentence}  & \textbf{Labelling} \\\hline
       \ctext[RGB]{173, 216, 230}{mm, oom: base root bonus on current usage} & Decision\\

\ctext[RGB]{255, 174, 66}{A 3\% of system memory bonus is sometimes too excessive in comparison to
other processes.} & Supporting Facts, Rationale\\

\ctext[RGB]{ 	255, 250, 205}{With commit a63d83f427fb (``oom: badness heuristic rewrite''), the OOM
killer tries to avoid killing privileged tasks by subtracting 3\% of
overall memory (system or cgroup) from their per-task consumption.} & Supporting Facts\\

\ctext[RGB]{255, 174, 66}{But
as a result, all root tasks that consume less than 3\% of overall memory are considered equal, and so it only takes 33+ privileged tasks pushing the system out of memory for the OOM killer to do something stupid and kill dhclient or other root-owned processes.} & Supporting Facts, Rationale\\

\ctext[RGB]{ 	255, 250, 205}{For example, on a 32G
machine it can't tell the difference between the 1M agetty and the 10G
fork bomb member.} & Supporting Facts\\

\ctext[RGB]{ 	255, 250, 205}{The changelog describes this 3\% boost as the equivalent to the global
overcommit limit being 3\% higher for privileged tasks, but this is not
the same as discounting 3\% of overall memory from *every privileged task
individually* during OOM selection.}  & Supporting Facts\\

\ctext[RGB]{173, 216, 230}{Replace the 3\% of system memory bonus with a 3\% of current memory usage
bonus.}  & Decision\\

\ctext[RGB]{200, 162, 200}{By giving root tasks a bonus that is proportional to their actual size,
they remain comparable even when relatively small.} & Rationale, Decision \\

\ctext[RGB]{244, 194, 194}{ In the example
above, the OOM killer will discount the 1M agetty's 256 badness points
down to 179, and the 10G fork bomb's 262144 points down to 183500 points
and make the right choice, instead of discounting both to 0 and killing
agetty because it's first in the task list.}  & Rationale\\
    \end{tabular}
    \caption{An example commit  with labelled sentences from our dataset}
    \label{tab:coloured_commit}
\end{table*}

\subsubsection{Dataset Insights}

An insight from this dataset is that there seems to be a trend for about 40-50\% of the sentences in a commit message to contain rationale information~\cite{FSE-SRC}. For example, four out of nine sentences in Table~\ref{tab:coloured_commit} contain rationale. This trend could indicate a ``natural'' amount of rationale that developers include in their commit messages, or maybe a guideline for developers to target. 

\begin{figure}[t]
    \centering
    \includegraphics[width=0.3\textwidth]{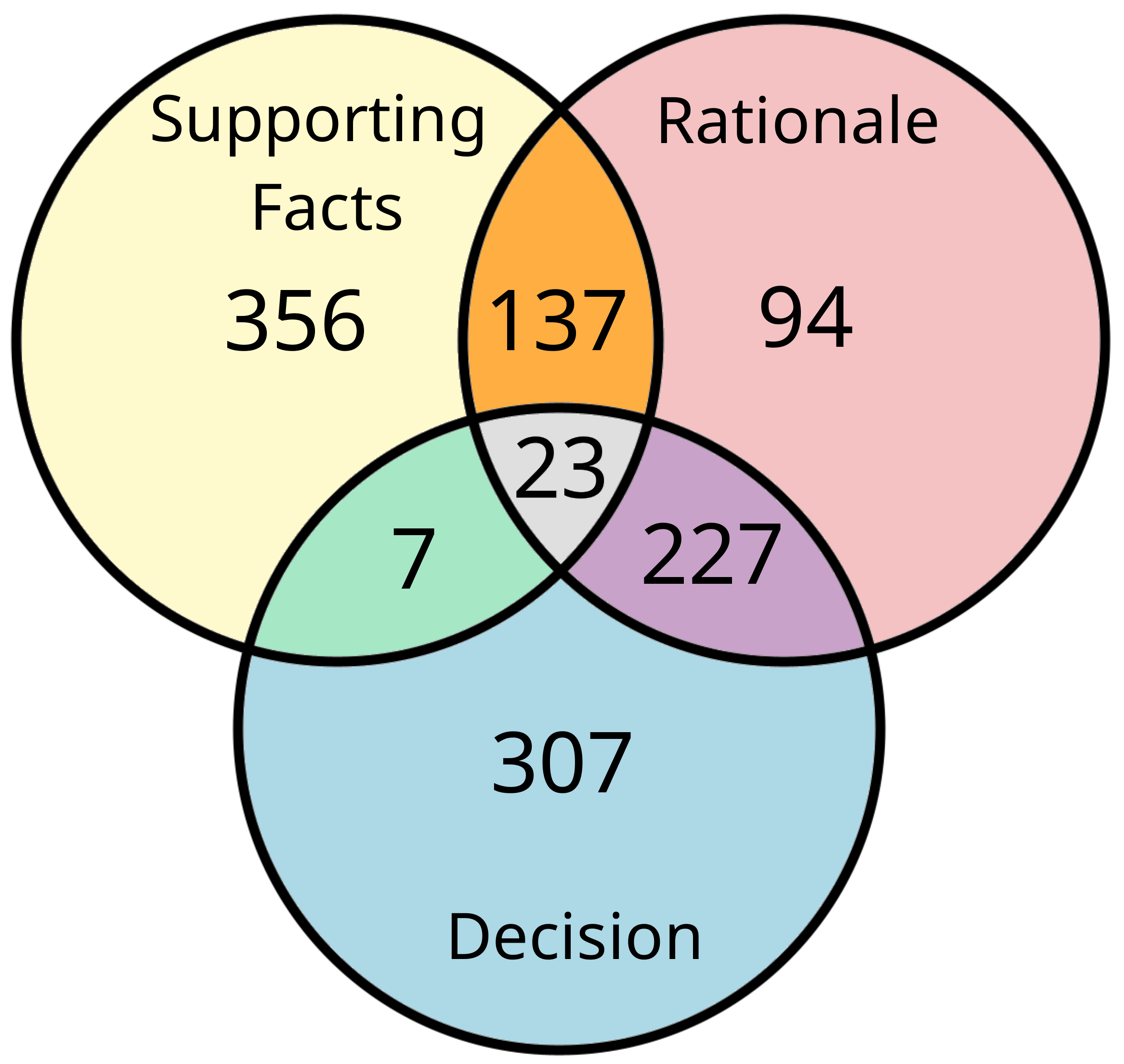}
    \caption{Distribution of the sentences in our OOM dataset}
    \label{fig:dataset_categories}
\end{figure}

Figure~\ref{fig:dataset_categories} shows the distribution of the sentences over the identified categories\footnote{Note that the colouring shown in Figure~\ref{fig:dataset_categories} for each category is also reflected throughout other tables and figures in this article.}.  This distribution motivates our creation of the \textit{supporting facts} category, as there are many sentences that contain supporting facts but do not contain rationale. Also evident is the clear separation between sentences containing \textit{decisions} and those with \textit{supporting facts}. However, the category of \textit{rationale} is less clear-cut, with a substantial overlap between sentences with rationale and the other categories. One interpretation is that this indicates the subjective nature of rationale. Another possibility is that \textit{rationale} is present in these sentences to motivate \textit{decisions} and to provide value judgments on the existing state of the system as discussed in \textit{supporting facts}. Examples of these multi-label sentences are found in Table~\ref{tab:coloured_commit}.


\section{Extracting Rationale Information}
\label{sec:extracting}

This section will discuss a key part of the Kantara framework: the classification of sentences with respect to whether they contain \textit{rationale}, \textit{decisions}, or \textit{supporting facts}. We discuss our approaches and provide an evaluation of our approach on our Linux subsystem dataset discussed in Section~\ref{sec:oom_labelling}.


Considering the subjective nature of the rationale information and the notable overlap between the different categories (Fig.~\ref{fig:dataset_categories}), we investigated two classification approaches: \textit{binary  classification} (we do not consider the overlap), and \textit{multi-label classification} (we consider the overlap). We selected the TF-IDF vectorizer to embed the 
commit messages. The vectors were then input
into various ML models.  We used the sklearn\footnote{\url{https://scikit-learn.org/stable/index.html}} library for the models implementation and kept the default parameters values. The models were trained and
evaluated using 10-fold cross-validation. That is, the data was randomly divided into ten equal splits, and nine of them were used for training and one for evaluating performance. We report the mean scores from these evaluations.

\begin{table*}[tp]
    \footnotesize
     \caption{Binary classification evaluation }
    \centering
    \begin{tabular}{|c|c|c|c|c|c|c|c|c|c|c|c|c|}
    \hline
      \multirow{2}{*}{\textbf{Model}}  & \multicolumn{4}{c|} {\textbf{Decision}} &  \multicolumn{4}{c|} {\textbf{Rationale}}& 
       \multicolumn{4}{c|} {\textbf{Supporting Facts}}\\
       \cline{2-13}
          & Accuracy & Precision & Recall & F1 & Accuracy & Precision & Recall & F1 & Accuracy & Precision & Recall & F1  \\
         \hline
         Logistic Regression & 0.72  & 0.75 & 0.20  & 0.30  & \textbf{0.69} & 0.20 & 0.03 & 0.05  & 0.70  & \textbf{0.75} &0.13  & 0.21  \\
          \hline
          Decision Tree & \textbf{0.78}  & 0.71  & \textbf{0.61} & \textbf{0.64} & 0.68 & \textbf{0.51} & \textbf{0.57} &\textbf{0.53} &\textbf{0.71} & 0.59  & \textbf{0.46} & \textbf{0.51} \\
           \hline
             
          SVM &0.73 & \textbf{0.85} &0.22 &0.33 & \textbf{0.69}  &  0.20  & 0.03 &  0.04& \textbf{0.71}   & \textbf{0.75}  & 0.17 & 0.27 \\
          \hline
    \end{tabular}
   
    \label{tab:binary_classification}
\end{table*}

\subsection{Binary classification}
In binary classification problems, any of the samples from the dataset takes only one label out of two classes. 
We therefore consider the subset of the OOM dataset with the sentences that were labelled with only one label. Specifically, we ended up with 307 decision sentences, 94 rationale sentences and 356 supporting facts sentences. Since the data is imbalanced, we apply under-sampling to the majority classes and consider only 100 decision sentences and 100 rationale sentences. We then trained different  classifiers considering one label at a time (i.e, when trying to classify decision sentences, rationale and supporting facts sentences we labelled as negative). 

Table~\ref{tab:binary_classification} reports the classification results of the widely-used  binary classification models: Logistic Regression, Decision Tree and Support Vector Machines (SVM)~\cite{wang2005support}. We report the Accuracy, Precision, Recall and F1-score evaluation metrics. 

From Table~\ref{tab:binary_classification}, we extract four insights. First, the overall recall across the binary classification techniques is low. Second, the Decision Tree algorithm gave the overall best results. Third, the classification of the \textit{Decision} sentences was more successful than the classification of the \textit{Supporting Facts} sentences. Finally, the performance of the three classifiers when classifying \textit{Rationale} sentences was rather low.









\subsection{Multi-label classification}
The multi-label classification is the most natural approach that applies to our problem as any sample from the dataset can be labelled with more than one label.  In this experiment, we considered 1151 sentences (distributed as shown in Fig.~\ref{fig:dataset_categories}) and we tried widely-used models for multi-label classification including the
eXtreme Gradient Boosting (XGBoost)~\cite{chen2015xgboost}. 
We report the micro-averaged evaluation results over the three categories in Table~\ref{tab:multi_label}. Results indicate that the XGBoost classifier gave the overall best performance. 

We split the dataset to 60\% train set and 30\% test set. We train XGBoost on the train set and test it on the test set. We report  its performance on the three categories in Table~\ref{tab:multi_label}. Results indicate that the best classification results were for the \textit{Decision} label, the second best were for the \textit{Supporting Facts}. The \textit{Rationale} classification results were the worst.

 \begin{table}[htbp]
\caption{Multi-label classification  evaluation}
\centering
\begin{tabular}{|c|c|c|c|c|}
\hline
\textbf{Model}& \textbf{Precision$^{\star}$} & \textbf{Recall$^{\star}$} & \textbf{F1 score$^{\star}$}   \\

\hline
Random Forest  &  \textbf{0.73} &  0.51  & 0.60  \\
\hline
XGBoost & 0.67 & \textbf{0.60} & \textbf{0.63} \\ 

\hline 
KNN & 0.60  & 0.50 & 0.54  \\

 

\hline
\multicolumn{4}{|l|}{$^{\star}$ Micro-averaged }\\

\hline
\multicolumn{4}{|l|}{\hfil \textbf{XGBoost classification evaluation}}\\
\hline
  \textbf{Label} & \textbf{Precision} &  \textbf{Recall} & \textbf{F1-score} \\
\hline

  Decision     &    \textbf{0.76}   &     \textbf{0.69}     &   \textbf{0.72} \\     
Rationale      &   0.62    &    0.41    &    0.49   \\    
Supporting Facts     &    0.64    &    0.68      &  0.66  \\
  \hline
\end{tabular}
\label{tab:multi_label}

\end{table}

\subsection{Summary}

These results indicate that the most challenging task is rationale classification, possibly due to the overlap of categories in the messages themselves, as seen in Figure~\ref{fig:dataset_categories}. We find that the \textit{Decision Tree} and \textit{XGBoost} techniques are most promising, but further investigation is needed to provide satisfactory results.

         
    



\section{Structuring Rationale Information}
\label{sec:structuring}

This section will discuss our contribution on \textit{structuring commit rationale information} in an ontological-based manner. 

\subsection{Decision/Rationale (DR) Graph}

In our previous work~\cite{dhaouadi2022end} we defined a graph structure to explicitly represent commits, their labelling, and their relationships. For example, we related \textit{conflicting} or \textit{similar} commits together and represented the \textit{rationale} for each commit as free text. Our current work is investigating the structure and classification of the rationale for one commit; we thus omit those inter-commit relationships here. Instead, from our insights from our labelling procedure (Section~\ref{sec:oom_labelling}), we now focus on rationale information at the \textit{sentence level} where sentences can have multiple labels.

\begin{figure}[tbh]
    \centering
    \includegraphics[width=0.485\textwidth]{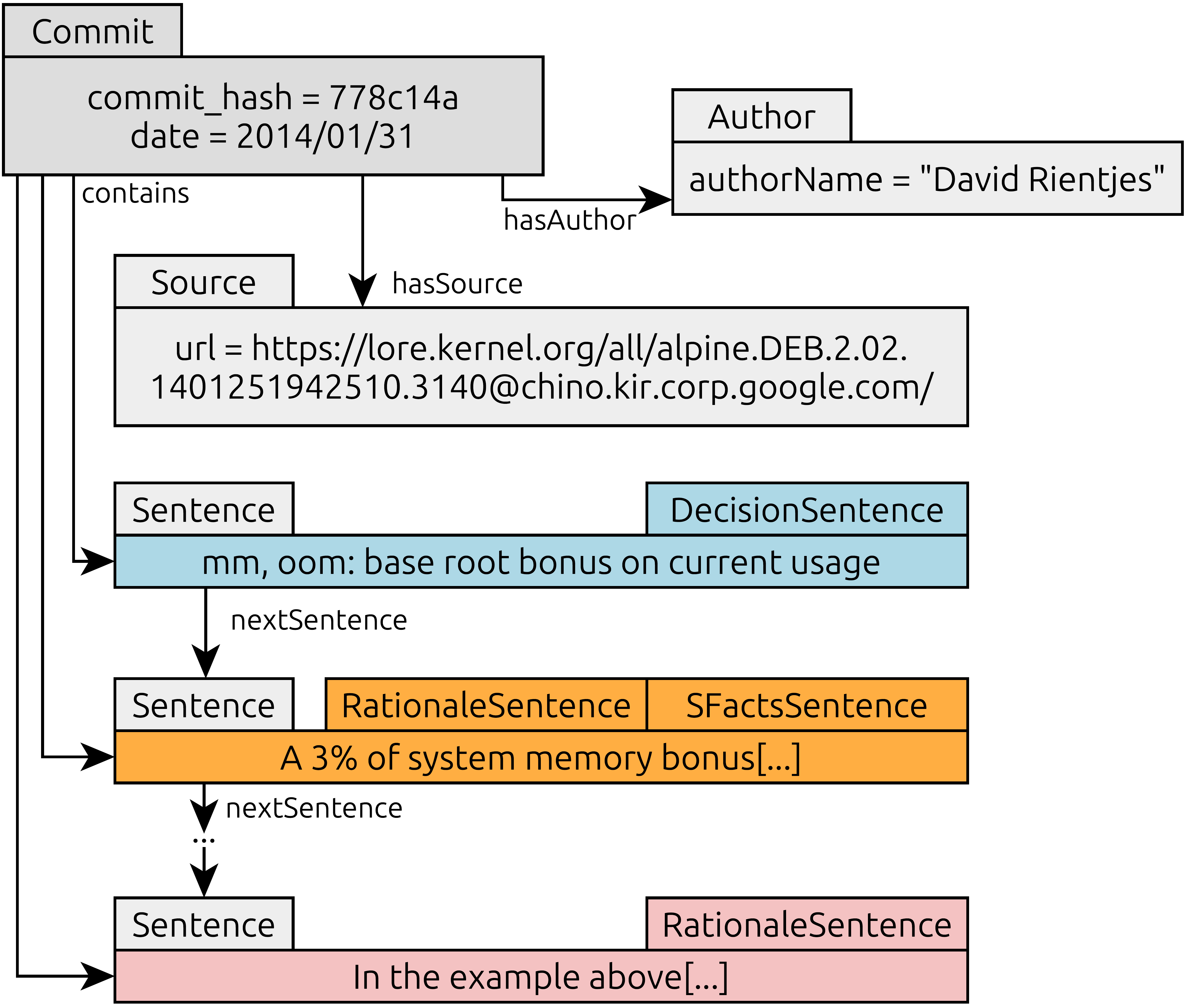}
    \caption{Decision/Rationale Graph}
    \label{fig:dr_graph}
\end{figure}

Table~\ref{tab:coloured_commit} shows the example commit in textual form with sentences colored according to their label, while Figure~\ref{fig:dr_graph} represents that commit as a DR graph. A commit has the usual information like the commit hash and the date, and each commit is linked to its author and the individual sentences within the commit text. Each \textit{Sentence} at the bottom of Figure~\ref{fig:dr_graph} has a link to the \textit{nextSentence} (to retain the structure of the commit), and are multi-typed with each label. Section~\ref{sec:inferencing} explains how these classifications are inferred automatically through ontological semantics, which aids the analysis and reporting functionality of the Kantara framework.


\subsection{Ontological Representation}
\label{sec:ontology}

The DR graph described in the previous section explicitly captures concepts and their relation. This makes it natural to opt for a representation of this DR graph using an \textit{ontological basis}, as ontologies represent ``a shared understanding of a domain''~\cite{uschold1998knowledge}.

\begin{figure*}
\centering
\begin{subfigure}{.48\textwidth}
\centering
\includegraphics[width=\textwidth]{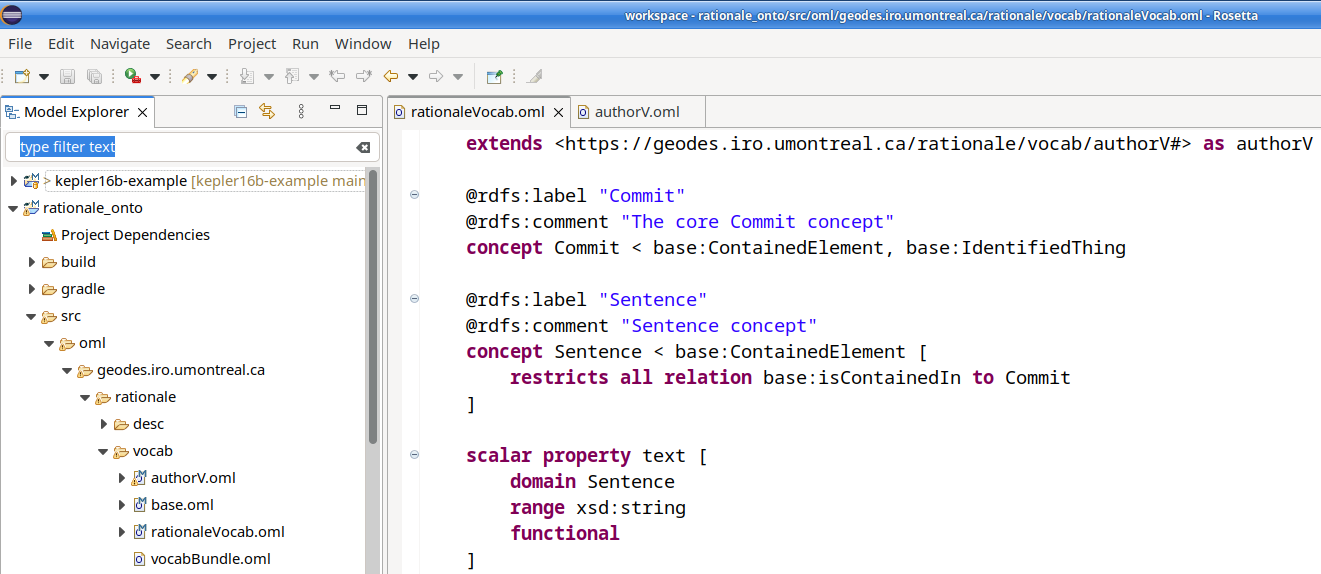}
\caption{Textural format}
\label{fig:rationaleV}
\end{subfigure}%
\begin{subfigure}{.48\textwidth}
\centering
\includegraphics[width=\textwidth]{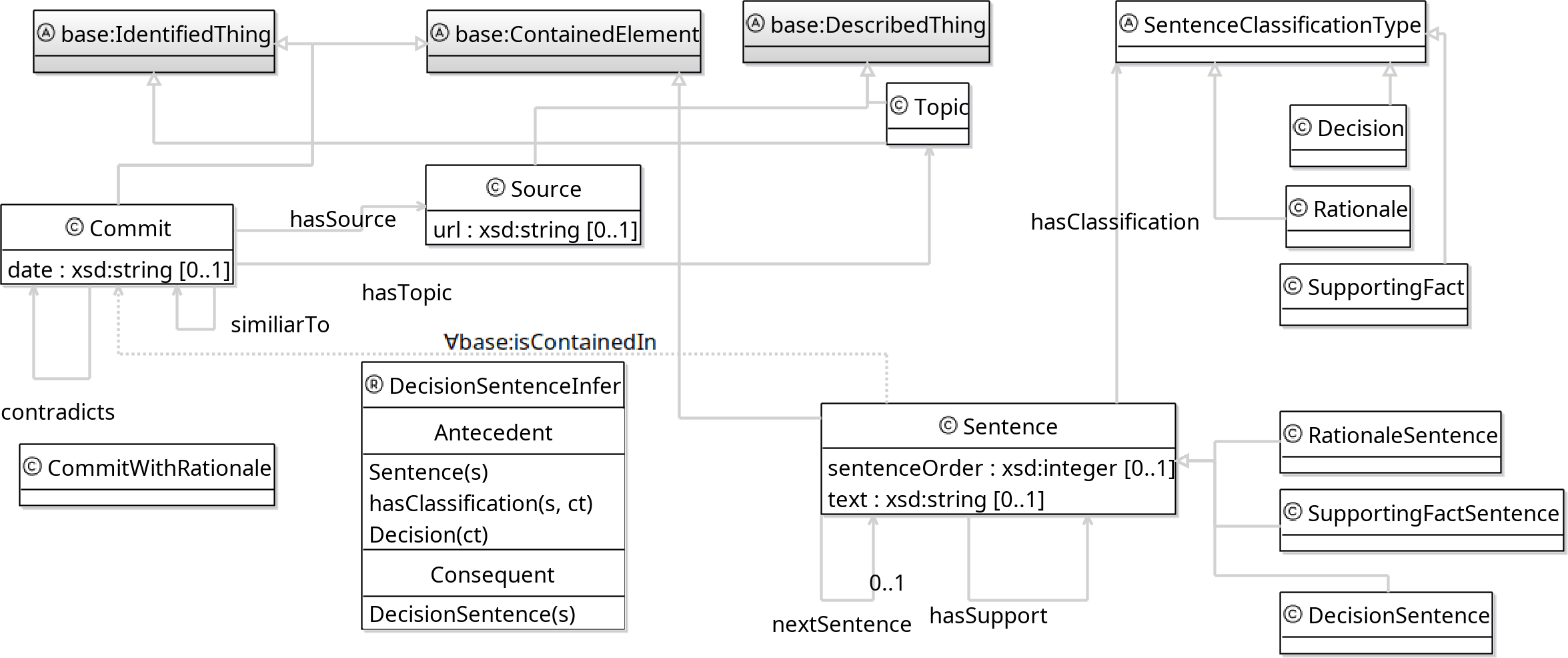}
\caption{Graphical format}\label{fig:rationaleV_graphical}
\end{subfigure}%
\caption{Rationale vocabulary editing within the Rosetta component of openCAESAR}
\label{fig:rosetta}
\end{figure*}

For the concrete ontological graph implementation, we use the openCAESAR framework\footnote{\url{http://www.opencaesar.io/}} developed by the NASA Jet Propulsion Laboratory (JPL) instead of a traditional ontological tool such as Proteg\'{e}. This is due to the approach of openCAESAR~\cite{Elaasar2023} which abstracts over the complex logical foundations of the well-known Web Ontology Language (OWL) to instead provide a high-level language for defining ontologies using the Ontological Modeling Language (OML).

Briefly, the openCAESAR methodology for creating ontologies focuses on the creation of a) \textit{vocabulary models} which correspond to ontology concepts (the A-box), and b) \textit{description models} which correspond to ontology individuals (the T-box). 
The Rosetta editor provided by the openCAESAR framework is shown in Figure~\ref{fig:rosetta}. In Figure~\ref{fig:rationaleV}, the textual OML syntax is shown for the \textit{rationale} vocabulary, defining the core concepts for \textit{Commits} and \textit{Sentences} and their relationship. Listing~\ref{lst:oml_rationale}) reproduces a portion of this vocabulary. Figure~\ref{fig:rationaleV_graphical} presents the OML graphical syntax. 

 \lstset{language=oml}
\begin{lstlisting}[
    breaklines=true,
    keepspaces=false,
    breakindent=0pt,
  % basicstyle=\ttfamily\footnotesize\scriptsize,
    basicstyle=\ttfamily\footnotesize,
%    numbers=left,
    caption={Excerpt of the OML rationale vocabulary model},
    label={lst:oml_rationale}]
aspect SentenceClassificationType
concept DecisionSentence < SentenceClassificationType
concept RationaleSentence < SentenceClassificationType
concept SupportingFactSentence < SentenceClassificationType
relation entity SentenceClassification [
    from Sentence
    to SentenceClassificationType
    forward hasClassification
    asymmetric
]

 \end{lstlisting}

On the right of Figure~\ref{fig:rationaleV_graphical} are the types of sentence classification and the subtypes of sentences which connect to each of these types of classification. We leverage the ontological approach to assign these subtypes to individual sentences through \textit{inferencing} discussed in Section~\ref{sec:inferencing}. 


The advantages are that a) subtypes of sentences can be automatically inferred by the ontological reasoner, and b) sentences can be labelled with multiple classification types. While this could be adequately captured using a standard meta-modelling approach, we found that this ontology-based approach simplified our analyses and more accurately captured the meaning of our labelling.

\subsection{Inferencing}
\label{sec:inferencing}

An advantage of storing the commit and rationale information in an ontology is that ontologies naturally form a graph structure (a ``knowledge graph'' or ``knowledge base'') suitable for inference and querying~\cite{oakes2021structuring}. We leverage this inferencing capability of ontologies to perform additional reasoning on our data set. From the Rosetta editor, tasks are available to run the Pellet ontological reasoner to perform this inferencing and check the consistency of the vocabulary and descriptions. 

An example of this reasoning in the rationale vocabulary is found in Listing~\ref{lst:oml_rules}. The \textit{rule} defines a antecedent/precedent \textit{rule} where if \textit{s} is a sentence and it has a classification of \textit{ct}, where that \textit{ct} is \textit{Rationale}, then the sentence \textit{s} should be inferred to be a \textit{RationaleSentence}. While this is a straightforward rule, the benefit is that it simplifies the further analysis and reporting of which sentences contain rationale. That is, the analysis and reporting steps will not have to follow extra edges to determine if a sentence is classified as having \textit{rationale} or not.

 \lstset{language=oml}
\begin{lstlisting}[
    breaklines=true,
    keepspaces=false,
    breakindent=0pt,
  % basicstyle=\ttfamily\footnotesize\scriptsize,
    basicstyle=\ttfamily\footnotesize,
%    numbers=left,
    caption={Classification rules for Sentences and Commits},
    label={lst:oml_rules}]

concept DecisionSentence < Sentence
concept RationaleSentence < Sentence
concept SupportingFactSentence < Sentence 

rule RationaleSentenceInfer [
    Sentence (s) & hasClassification (s, ct) & Rationale (ct) -> RationaleSentence (s)
]

concept CommitWithRationale = Commit [
    restricts base:contains to min 1 RationaleSentence
]
 \end{lstlisting}

The \textit{restricts} inference rule in Listing~\ref{lst:oml_rules} goes farther with a similar typing action. Here, a commit that \textit{contains} at least one \textit{RationaleSentence} is a \textit{CommitWithRationale}. Again, this simplifies analysis and reporting by removing the need to resolve links during those phases. For instance, without this inferencing, each analysis and report would have to determine if a commit contains a sentence with a classification type of rationale. Instead, this inferencing allows us to directly query for whether the commit is a \textit{CommitWithRationale}.

Section~\ref{sec:analysis} presents an analysis query and visualization which we have developed within the openCAESAR framework to better understand the rationale information. This query directly relies on these inferences.
 
\section{Analysis}
\label{sec:analysis}

This section discusses the last step of the Kantara framework: analysis and visualization of the extracted rationale information. The intention of Kantara is to provide the user with insights into the presence of rationale in the commit sentences. This can be used to increase the quality of commit messages. For example, the patch submission process could require a certain percentage of the commit to contain rationale. Or, this could be used to identify authors or subsystems with insufficient rationale, so that interventions could be made.

\subsection{Graph Queries}

We employ the openCAESAR framework to build the rationale ontology, such that
Kantara can produce OML \textit{description} models through text generation which conform to this vocabulary (Section~\ref{sec:ontology}). Then, we can execute SPARQL queries on these description models to obtain results. For a developer to better understand the rationale information present in our DR graph, we have created queries such as querying for a list of authors and their commits, and querying for those sentences classified as \textit{DecisionSentence}, \textit{RationaleSentence}, or \textit{SupportingFactSentence}.

Listing~\ref{lst:query} demonstrates a more comprehensive SPARQL query (with omitted PREFIXes) which can be executed within the openCAESAR framework to return a JSON result containing  commits, their authors, and whether each commit contains rationale. This table also contains the text and classification type(s) for each sentence in that commit.

Note that this query directly relies on the inference rules discussed in Section~\ref{sec:inferencing}. With the typing provided by the inference rules, the query can be more concise. This motivates our choice of using an ontological approach for storing this rationale information.

 \lstset{language=SPARQL}
\begin{lstlisting}[
    float,
    breaklines=true,
    keepspaces=false,
    breakindent=0pt,
  % basicstyle=\ttfamily\footnotesize\scriptsize,
    basicstyle=\ttfamily\footnotesize,
%    numbers=left,
    caption={Query for commits and labelled sentences},
    label={lst:query}]
SELECT ?author ?commit_id ?order ?text 
(BOUND(?hasRationale) AS ?isCommitWithRationale)
(BOUND(?rct) AS ?isSentenceRationale) 
(BOUND(?dct) AS ?isSentenceDecision) 
(BOUND(?sct) AS ?isSentenceSupporting)
WHERE {
  ?commit a rationale:Commit .
  ?commit baseV:hasIdentifier ?commit_id .
  ?commit rationale:hasAuthor ?author_id .
  ?author_id authorV:authorName ?author .
  ?commit baseV:contains ?s .
  ?s rationale:text ?text .
  ?s rationale:sentenceOrder ?order
  OPTIONAL {
  	?commit ?hasRationale rationale:CommitWithRationale .
	}
	OPTIONAL {
  	?s ?rct rationale:RationaleSentence .
	}
	OPTIONAL {
  	?s ?dct rationale:DecisionSentence .
	}
	OPTIONAL {
  	?s ?sct rationale:SupportingFactSentence .
	}
}
ORDER BY ?commit_id ?order
 \end{lstlisting}


\subsection{Visualization}

\begin{figure*}[tbh]
    \centering
    \includegraphics[width=0.95\textwidth]{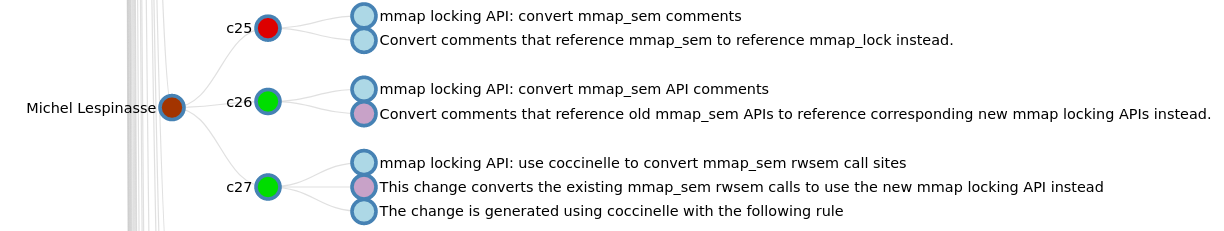}
    \caption{Interactive visualization of authors, their commits, and commit sentence text}
    \label{fig:commit_viz}
\end{figure*}

The results of the query in Listing~\ref{lst:query} can be inspected in tabular form but we developed Kantara to also support interactive visualization, e.g., to display authors, their commits, and the sentences within those commits. The intention is for developers to use such visualizations to understand the rationale information present in each commit.

Figure~\ref{fig:commit_viz} shows a screenshot of a particular visualization that that centres committers' practices. In the example shown we focus on OOM developer \textit{Michel Lespinasse}, who has authored one commit (\textit{c25}) without rationale, and two commits (\textit{c26}, \textit{c27}) with rationale. The interactivity of the visualization allows the user to collapse or expand the children of nodes to focus on certain authors or commits. The visualization runs within a web browser and is built using Javascript and the D3 library, modified from an example provided in openCAESAR.

Authors are the left-most node in Figure~\ref{fig:commit_viz} where the color indicates whether their commits contain rationale or not. Each commit (\textit{c25}, \textit{c26}, etc.) are shown to the immediate right. Finally, the sentences making up that commit are on the right-hand side, with the same coloring as in Figure~\ref{fig:dataset_categories} and Table~\ref{tab:coloured_commit} to indicate the classification of the sentence. In Figure~\ref{fig:commit_viz}, all seven sentences are labelled as \textit{decisions}, and two are also labelled as \textit{rationale}.

This visualization provides insight in committers' rationale documentation practices in these commits. It lets a developer understand the presence of decisions, rationale, and supporting facts written by others, exposing commonalities in a set of commits. It also highlights committers whose commits do not provide rationale. This could help the community to improve its rationale documentation practices, by alerting maintainers to commits with insufficient rationale information.





\section{Related Work}
\label{sec:related_work}


We divide related work into two categories: \textit{extraction of rationale} and \textit{representations of rationale}.

\subsection{Extracting Rationale}

Sharma \textit{et al.} employed  a heuristics-based approach  to try to unearth rationale from  Python Email Archives~\cite{sharma2021extracting}.  
In \cite{mccall2019using}, the authors propose ASGAR, a semantic grammar-based approach to automatically capture and structure design rationale. These works differ from ours as we propose an NLP-based approach.

Other researchers have employed machine learning (ML) techniques to extract rationale. In~\cite{li2020automatic}, the authors tried to automatically identify decisions from textual artifacts. They created a labelled ground
truth from the Hibernate developer mailing list and tried various classification models and
configurations. Similarly, in~\cite{licommit2023}, the authors  analyzed  Apache commit messages to study the impact of message quality on software defect proneness. As a quality indicator, the authors considered the binary presence of rationale information and they manually labelled the commits as containing or not motivation/rationale information. These  works employ binary classification, thus our work differs by proposing a rationale representation with different multi-classifications (Table~\ref{tab:codebook}).

ML techniques have been employed in an attempt to solve the \textit{capture} problem (i.e., the large effort required to capture the rationale manually and adhere to a specific representation)~\cite{burge2008design}. They aim at bridging the gap with the unused existing rationale representation models such as Burge's rationale model~\cite{burge2008rationale}. For example, in~\cite{alkadhi2017react},  the authors propose annotating chat messages that contain rationale by capturing five rationale
elements adapted from the IBIS model~\cite{rittel1989issue}. 
In~\cite{liang2012learning}, the authors focus on designing an algorithm to extract and structure  design rationale from design documents based on the 
ISAL model~\cite{liu2010new}.
Rogers \textit{et al.} investigated the usage of ontology and linguistic features to identifying rationale from Chrome bug reports~\cite{rogers2015using}, 
Lester \textit{et al.} investigate evolutionary algorithms for optimal features to extract and classify design
rationale from Chrome bug reports and design discussion transcripts~\cite{lester2019identifying}, and Mathur tried to improve Lester's classification results for the rationale with new features and algorithms~\cite{mathur2015improving}.  These model-based classification approaches differ from the  data-driven categorization we employ in this work.

\subsection{Representing Rationale}

Recently,  other  rationale representations have been proposed. AlSafwan \textit{et al.} proposed a model with 15 categories to represent rationale in code commits after conducting interviews and a survey~\cite{alsafwanDevelopersNeedRationale2022}. Hesse \textit{et al.} proposed a documentation model for decision knowledge built upon the 
results investigating the comments to 260 issue reports from the Firefox project~\cite{hesse2020supporting}.   Kleebaum \textit{et al.} built  upon Hesse’s documentation model and proposed Condec tools to support requirements engineers in documenting and exploiting decision knowledge for change impact analysis~\cite{kleebaum2020continuous}.  ConDec tools build up and visualize a knowledge graph consisting of knowledge elements and trace links. Soliman \textit{et al.} worked on an empirically-grounded ontology for architecture knowledge from StackOverflow~\cite{soliman2017developing}. Their work also supports automating architecture knowledge capturing and proposing solutions to effectively search for relevant architectural information in developer communities.
Bhat \textit{et al.} proposed AdeX, a framework to  extract and reuse architecture knowledge and to recommend alternative architectural solutions that could be considered during architectural design making~\cite{mahabaleshwar2020tool}. The authors propose a static and a dynamic architectural knowledge models.  In AdeX, the rationale of a decision refers to the  quality attribute the decision addresses. AdeX automatically identifies architectural elements in design decisions. These architectural elements are identified using concepts
captured in a publicly available cross-domain ontology.

The related work differs from our in two main ways. First, none of these works focus on  representing and managing the rationale specifically for the Linux kernel. Second, we provide a knowledge graph-based approach and pipeline using the state-of-the-art openCAESAR tool for constructing and querying ontologies.

 
\section{Conclusion and Future Work}
\label{sec:conclusion}

This article has presented the Kantara framework for structuring and extracting the rationale found in the commits of developers. We have presented our new classification of rationale information, focusing at the sentence-level and now including \textit{rationale}, \textit{decision}, and \textit{supporting fact} labels. We discussed approaches to classify each sentence using NLP approaches and a brief evaluation. Finally, we described how this rationale information is represented using an ontologically-based knowledge graph in the openCAESAR framework\cite{Elaasar2023}, along with the analysis and reporting capabilities in Kantara.

\subsection*{Challenges and Future Work}



\paragraph{Challenge 1: Expanding the Rationale Ontology}

Our categorization of components of the rationale of commits was data-driven where we came to a consensus after several discussions during labelling~\ref{sec:oom_labelling}. Our model here focuses on sentence-level labelling to evaluate our  Kantara framework and the NLP classification. In the future, we plan to extend our representation to include all the components of the rationale of  commit messages (e.g. \textit{Goal}, \textit{Need}, \textit{Benefit}, etc.~\cite{alsafwanDevelopersNeedRationale2022}) and other previous rationale representations (Section~\ref{sec:related_work}).

\paragraph{Challenge 2: Subjective labeling}
It is possible that our manual labelling
process has introduced unintentional bias. To address
this, the three authors labelled independently.  A
Fleiss kappa of 0.69  indicates a high reliability of our labelling. In the future, we are exploring avenues such as: a) including kernel developers themselves to validate our labels, or b) determining if large language models (LLMs) can provide suitable performance for this labelling task.

\paragraph{Challenge 3: Classification performance}

As shown in Section~\ref{sec:extracting}, the performance of the classification approaches was rather average. While it may not be possible to achieve perfect performance due to the subjective nature of this classification, we are attempting to improve these metrics. Concrete steps include a) applying hyper-parameter
tuning to all classifiers to find optimal hyper-parameters, b) expanding our labelled data set of commit messages, and c) investigating more advanced classification architectures such as Bidirectional Long Short-Term Memory (Bi-LSTM)~\cite{zhou2016attention} or pre-traineed language models, e.g., BERT~\cite{devlin2018bert}



\paragraph{Challenge 4: Generality}
The last challenge we wish to address is the threat to validity of generalizing this classification approach and our insights. Section~\ref{sec:example} describes how the  Linux kernel and the OOM subsystem have a particular development culture which emphasizes detailed commit messages. This culture may not be present in other projects, such that developers are not encouraged or required to produce commits with such detailed rationale information. In future work, we will investigate this by applying our Kantara approach to other codebases.




\section*{Acknowledgment}
This research work is partially funded by the Fonds de Recherche du Québec – Nature et Technologies (FRQNT) Doctoral Research Scholarship (B2X).

\bibliographystyle{IEEEtran}
\bibliography{rationale.bib, ml.bib}

\end{document}